\documentclass[prb,twocolumn,showpacs,amsmath,superscriptaddress,amssymb]{revtex4}

\usepackage{graphicx}
\usepackage{dcolumn}
\usepackage{bm}

\begin{document}


\title{Semiclassical Lattice effects on interband tunneling of a two-state system}
\author{Ryuji Takahashi}
\email{ryuji.takahashi@riken.jp}
\affiliation{Condensed Matter Theory Laboratory, RIKEN, Wako, Saitama 351-0198,  Japan}
\author{Naoyuki Sugimoto}
\affiliation{Department of Applied Physics, The University of Tokyo, Japan}
\date{\today}

\begin{abstract}
Previously, we have shown that the transition probability of the Landau-Zener problem in periodic lattice systems becomes large by taking into account the nonlinearity of the energy spectra, compared with the probability by the conventional Landau-Zener formula.
The enhancement comes from the nonlinearity peculiar to the periodic lattice system,
 and this effect from the lattice on transition action cannot be neglected in the transition process.  
 In the present paper, we first give a brief review of the previous work, and construct the transfer matrix of the Landau-Zener problem by the semiclassical description for lattice systems.
 Next, we study a ladder lattice system and show that the transition action obtains a phase due to the nonlinearity.
 Then, we consider the double-passage problem of the ladder system within the semiclassical description.
 We find the oscillation of the probability by the transition phase by the lattice effect.
This phase comes from the semiclassical analysis unlike the Stokes phase, and we show that the oscillation is mainly contributed by the transition phase by the lattice effect, when the hybridization of the ladder is strong.
\end{abstract}

\maketitle

\section{Introduction}
%
The Landau-Zener (LZ) problem gives a simple description for the quantum tunneling phenomena of two-level systems~\cite{Landau32,LandauB,Zener32}, and this tunneling problem has been studied in a variety of fields, since the quantum two-level systems are ubiquitously obtained. 
In addition to the nonadiabatic transition, interference of the tunneling amplitude has been studied for periodic systems~\cite{NikitinB,NakamuraB,Shimshoni91, Nori10,Fuchs12, Suominen12}.
In the semiclassical description, the nonadiabatic transition occurs on a band edge,
 and the interference 
 appears between the amplitudes coming from different band edges with the phase difference~\cite{Leggett95,Montambauxh15,Li15}. 
Then, the oscillation of the transition probability occurs, and this interference phenomenon is often called St\"{u}ckelberg interferometry \cite{Tarruell12,Stueckelberg32, Mark07}.
Since this phenomenon is expected to provide information of the two-state system, it has been studied in various systems, e.g., atomic collisions~\cite{NikitinB}, and semiconductor superlattices~\cite{Rotvig95, Rotvig96}.

The calculation of the transition probability is done by the LZ formula in the conventional LZ problem. However, this formula lacks the nonlinear effect coming from the periodicity of the lattice. 
According to our recent work~\cite{Takahashi17}, the transition phenomena of the two-level system of the bulk periodic lattice have been studied by using the instanton method based on the path integral for the Bloch states \cite{Takahashi17}, and the enhancement of the probability have been shown, compared with that by the LZ formula.
 We call the correction of the tunneling probability by the lattice the lattice effect, which comes from the band structure of the material.
While interesting phenomena have been shown in large values of the transition probability in earlier studies \cite{Tarruell12,Stueckelberg32, Mark07,Montambauxh15,Li15}, we have shown that the lattice effect becomes prominent when the probability is much smaller than unity by using a dimerized one-dimensional (1D) chain.
 The enhancement of the probability comes from the periodicity of the energy spectra, and the tunneling probability is small in insulating materials.
Then, the lattice effect is expected to give a drastic enhancement of the tunneling probability for general insulators.
The instanton method gives a simple description for the periodic lattice, and one can obtain the nonlinear effect without the solution of the Schr$\ddot{\mathrm{o}}$dinger equation.
Then, this method is expected to be applicable to various  transition phenomena in lattice systems.
Moreover,  recently studies of the metal-insulator transition of materials have been done \cite{Asamitsu97,Miyano97}. 
In these materials, the transition is described by the 1D chain of the LZ problem.
The tunnel current is contributed by the transition of electrons on the 1D chains, and the number of these chains is expected to be large in many cases , because the tunneling current has been detected despite of the smallness of the transition probability. 
 Therefore, the hybridization between 1D chains is expected, and it may causes a change in the energy spectra of the system.
Then, the change is expected to affect on the tunneling probability according to our previous work.

In this paper, we investigate the tunneling phenomenon of a ladder system; two parallel 1D lattice systems interact each other. 
We first study the transfer matrix for two-state systems with a constant gap by the instanton method.
Physically, this gap appears by the dimerization of quantum wires, for example.
Then, we show that the transfer matrix becomes unitary by the Berry phase term of the semiclassical Lagrangian, by calculating the both of the instanton and anti-instanton processes;  the conservation of probability of the tunneling current is shown by the instanton method.
 Next we investigate the transition probability in a ladder system by using the transfer matrix of the instanton.
 We find that the imaginary part of the transition amplitude appears due to the hybridization by the semiclassical method unlike the quantum phase\cite{Nori10, Fuchs12}, and this imaginary part affects the interference of the tunneling phenomena of the double-passage process.
 Furthermore, we show that this imaginary part by the lattice effect becomes dominant for the oscillation of the tunneling probability coming from the interference.
 The strength of the hybridization can be controlled, and we also discuss the observation scheme of the oscillation in materials. 
 
This paper is organized as follows. In Sec.~\ref{sec:sct}, we discuss the transfer matrix of the tunneling of the LZ problem for the semiclassical description.
In Sec.~\ref{sec:sshl}, we consider the transition phenomena for a ladder model and formulate the tunneling problem by using the method in Sec.~\ref{sec:sct}. In Sec.\ref{sec:prob}, we show the calculation results for the transition probabilities. 
\begin{figure}[htbp]
 \begin{center}
 \includegraphics[width=80mm]{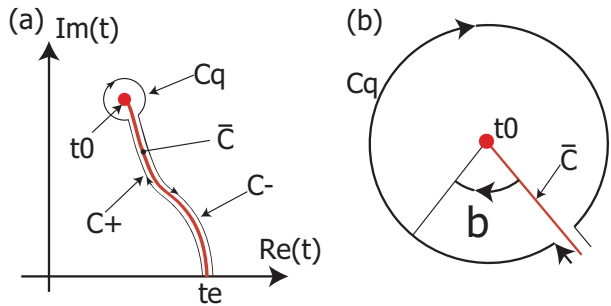}
 \caption{Schematic of the transition process $C$ consisting of $C_{\pm}$ and $C_{q}$.
(a) The initial and final states are on the band edge $k_{e}\equiv Ft_{e}$, at $t_{i}=t_{e} -\epsilon$ and $t_{f}=t_{e} +\epsilon$ with the infinitesimal time $\epsilon$.
 The bold curve ($\bar{C}$) from $t_{e}$ is given by Re$[E(k=Ft)]=0$ and is the branch cut.
$t_{0}$ is the branch point for the energy change $\xi\to-\xi$, and along the thin contour $C$, transition occurs.
$C_{\pm}$ are the contours along the branch cut $\bar{C}$ , and $C_{q}$ goes around the branch point with infinitesimal radius.
(b) Magnified schematic of $C_{q}$. $b$ is the angle around the branch point.
}
\label{fig:timecontour}
\end{center}\end{figure}
\section{Calculation method of the transfer matrix by the instanton of the Bloch states}\label{sec:sct}
In this section, we construct the transfer matrix of the LZ problem of the lattice system based on the instanton method of the Bloch states\cite{Takahashi17}. The transfer matrix of the tunneling $\mathcal{T}$ is defined by a $2\times 2$ matrix as \cite{Nori10}
 \begin{eqnarray}
\psi^{f}
&=&\mathcal{T}
\psi^{i},\\
\mathcal{T}&=&
\begin{pmatrix}
T'& R \\
R' &T
\end{pmatrix}.\label{eq:gentmat}
\end{eqnarray}
where $\psi^{f(i)}=(\psi_{u}^{f(i)}, \psi_{l}^{f(i)})^{t}$ is the wavefunction at the final (incident) state consisting of the upper and lower components.
$T(T')$ is the transition amplitude of the lower (upper) to the upper (lower) energy level: $|T|= |T'|$.
The energy transition $-\xi\to\xi$ occurs in the same band in the expression of the transfer matrix; the transition to the higher energy level is described by $\psi^{in}=(0,1)^t$ $ \to\psi^{out} =(0,1)^{t} $.
When the reflection coefficients have the relation, $R'=-R^{*}$, the unitarity of the transition matrix is satisfied. 
We obtain the transition coefficient by the instanton method, and the reflection coefficient is determined by the conservation law $|T|^2 +|R|^2=1$. In addition, the unitarity is given by the Berry phase term as shown later.

\subsection{Review of the transition for a two-state system}\label{subsec:RTM}
To calculate the transmission coefficient $T$, we consider the general one-particle Bloch Hamiltonian $H_{k}$ with the wavevector $k$. $H_{k}$ is diagonalized as $\hat{\xi}_{k}=\mathrm{diag}(\xi_{1k},\xi_{2k},\dots,\xi_{Mk})$, where $M$ is the number of bands.
According to the path integral procedure\cite{FeynmanB}, we consider time transition amplitude from $t=t_{i}$ to $t_{f}$.
In the presence with the constant external force $F>0$,
the transition amplitude by the path integral is given as \cite{Takahashi17}
 \begin{eqnarray}
\mathcal{K}(\Upsilon _{f},\Upsilon _{i})
&=& \int\mathcal{D}\Upsilon\exp\left[{i\int_{t_{i}}^{t_{f}}\mathrm{d}t~L }\right]
\label{eq:pathinteg2}
\end{eqnarray}
where $L$ is the Lagrangian, $\mathcal{D}\Upsilon =\mathcal{D}\eta^{\dagger}\mathcal{D}\eta $, and $k_{t}=F t+$const..
The spinor $\eta_{t}$ satisfies $\eta^{\mu\dagger}_{t} \cdot\eta_{t}^{\nu} =\delta_{\mu\nu}$, $^\forall j\in [1,\dots,N] $
for $\mu,\nu \in[1,\dots,M]$.
 In this study, we consider two-state systems $M=2$, and by using the effective Lagrangian, we obtain the transition amplitude with the semiclassical description.

We consider a 1D two-state system with the energies $+E(k)$ and $-E(k)$ labeled with the wavenumber $k$.
In the Brillouin zone, $E(k)$ is a periodic function with a finite width of the energy and becomes zero in some wavenumbers. In addition, we assume a constant interaction $\Delta$ between the two states. 
 Then, the Bloch Hamiltonian is represented as
 \begin{eqnarray}
H_{k}=\sigma_{z}E(k) +\sigma_{x}\Delta,\label{eq:Hamilt}
\end{eqnarray} 
where $\boldsymbol{\sigma}$ are the Pauli matrices for the band. The eigenvalues are $\xi_{\pm}=\pm\xi=\pm\sqrt{E(k)^2+\Delta^2}$.
The band edge appears for $E(k)=0$, and the nonadiabatic transition occurs there.
The action and effective Lagrangian are given as
 \begin{eqnarray}
S&=&\int_{t_{i}}^{t_{f}}\mathrm{d}t~L,\nonumber\\
 L 
&=&-i\dot{\eta}^{\dagger}\eta+F\frac{a_{k}}{2}\langle \sigma_{y}\rangle-\xi(k)\langle \sigma_{z}\rangle,
\label{eq:lag}
\end{eqnarray}
where $k=Ft$, $a_{k}=\frac{\Delta }{\xi^2}\partial_{k}E(k)$ is the Berry connection, and $\langle \boldsymbol{\sigma}\rangle\equiv\eta^{\dagger}_{t} \boldsymbol{\sigma} \eta_{t}$.
By the Euler-Lagrangian equation \cite{Xiao10}, we have simple solutions:
 \begin{eqnarray}
\cos\theta=\pm1,\ 
\varphi=n \pi,
\label{eq:ELs}
\end{eqnarray}
where we expressed $\langle \boldsymbol{\sigma}\rangle =(\cos\varphi \sin\theta,\sin\varphi \sin\theta,\cos\theta )$, $n$ is an integer, and we use $\varphi=0$ as the initial state in the following discussion, without loss of generality. 
In the previous work\cite{Takahashi17}, the semiclassical action is extracted by the solution (\ref{eq:ELs}), and 
the energy transition $\xi_{k}\to -\xi_{k}$ is described by the instanton method.
In this study, we consider a contour $C$ (Fig.~\ref{fig:timecontour}) on the complex plane for the instanton process, and we separate $C$ into three parts, $C_{+}$, $C_{-}$ and $C_{q}$ by assuming the presence of the branch point $\xi_{k}=0$ (Fig.~\ref{fig:timecontour}). 
The contour $C_{+(-) }$ goes to (from) the branch point $t = t_{0}$
 from (to) the initial point $t_{e}$ along the branch cut $\bar{C}$ given by Re$[E(k)]=0$, and $C_{q}$ goes around the branch point $t_{0}$.
The boundary condition of the action is given as $t_{i}=t_{e}-\epsilon$ and $t_{f}=t_{e}+\epsilon$ with the infinitesimal time $\epsilon$. 
Then, the instanton action is represented as
 \begin{eqnarray}
S_{\mathrm{in}}
&
=&\int_{C_{+}}\mathrm{d}t~L_{\mathrm{cl}}+ \int_{C_{-}}\mathrm{d}t~L_{\mathrm{cl}} +\int_{C_{q}}\mathrm{d}t L_{q}\label{eq:eflag}
\end{eqnarray}
with the semiclassical Lagrangian $L_{\mathrm{cl}}=-\xi \langle \sigma_{z}(t)\rangle$ on $C_{\pm}$.
 $L_{q}$ is given by the semiclassical solution on $C_{q}$.

 The transition amplitude is given as
 \begin{eqnarray}
\langle \eta_{f} |\mathrm{e}^{-i\int_{C}H\mathrm{d}t}| \eta_{i}\rangle=\mathrm{e}^{iS_{\mathrm{in}}}. \label{eq:pinst}
\end{eqnarray}
 $S_{\mathrm{in}}$ is calculated by the semiclassical solutions of the spinor $\eta$ on the respective paths $C_{\pm}$, $C_{q}$, and connect them continuously.
According to the solution (\ref{eq:ELs}), $\langle \sigma_{z}(t)\rangle$ is constant on the contours $C_{\pm}$, and we have two types of the transition process;
$\langle \sigma_{z}(t_{i})\rangle=\langle \sigma_{z}(t_{f})\rangle$ and $\langle \sigma_{z}(t_{i})\rangle =-\langle \sigma_{z}(t_{f})\rangle$.
When the band index is unchanged $\langle \sigma_{z}(t_{i})\rangle=\langle \sigma_{z}(t_{f})\rangle$, the energy transition is described, and this solution gives the transmission amplitude of the tunneling.
In addition, since $\langle \sigma_{z}(t)\rangle$ is constant throughout the energy transition process with $|\langle \sigma_{z}(t)\rangle|=1$,
 we have $\langle \sigma_{y}(t)\rangle = 0$, and the effective Lagrangian on the branch point is given as $L_{q}=L_{\mathrm{cl}}=0$.
Then, the transition amplitude is written as
 \begin{eqnarray}
T&=&\mathrm{e}^{iS_{\mathrm{cl}}},\nonumber\\
S_{\mathrm{cl}}&=&-2\kappa \int_{\bar{C}} \mathrm{d}t~ \xi
\label{eq:traam}
\end{eqnarray}
where we express the constant $\kappa = \langle \sigma_{z}\rangle$ for simplicity, and the dilute instanton gas approximation $P=|T|^2\ll1$ is used.
As shown in the following results, the hallmark of our theory can be seen for $|T|^2\ll1$.

\subsection{Transfer matrix for a two-state system}\label{subsec:TM}
When the band transition occurs $\langle \sigma_{z}(t_{i})\rangle =-\langle \sigma_{z}(t_{f})\rangle$, 
the first two terms of the semiclassical action (\ref{eq:eflag}) cancel each other, since the energy transition $\xi\to-\xi$ occurs at the branch point.
This solution describes the reflection process, and the reflection phase by the action on $C_{q}$ appears.
 Here, to obtain the reflection phase, we assume that 
 $E(k=Ft)$ is expressed as a purely imaginary function $E=i\Tilde{E}$ by imaginary time $t=x+iy$ on the branch cut $\bar{C}$.
Then, we obtain $\bar{C}$ by the equation Re$[E(x+iy)]=0$ on the complex plane, and the branch point is given by $\Delta^2-\Tilde{E}^2=0$.
Next, we denote the branch point as $t_{0}=x_{0}+iy_{0}$, and in the vicinity of $t_{0}$, i.e., on $C_{q}$,
 we have the magnitude of the energy $\xi =\sqrt{\Delta^2+[E(t_{0}+\delta t) ]^2} \sim\sqrt{2 E(t_{0}) \mathrm{d}_{t}E(t_{0})\delta t} \to 0$ for $|\delta t|\to 0$.
Then, the Lagrangian on $C_{q}$ is expressed as
 \begin{eqnarray}
 L _{q}
&\sim&-i\dot{\eta}^{\dagger}_{q}\eta_{q}-i\lambda_{E} \frac{1}{4\delta t}\langle \sigma_{y}\rangle,\label{eq;lagq}
\end{eqnarray}
 where 
the spinor on $C_{q}$ is expressed as $\eta_{q}$, and $\lambda_{E}=\pm1$ corresponds to the sign by $\Tilde{E}(t_{0})=\lambda_{E} \Delta$.
For the Berry phase, the change of the physical parameters act only on the sign in this model\cite{Davis76}.
By equation of motion, we have
 \begin{eqnarray}
\dot{\eta}_{q}=\frac{\lambda_{E}}{4\delta t} \sigma_{y} \eta_{q}.
\end{eqnarray}
 By using the above form, the phase shift is simply obtained as follows.
As shown in Fig.~\ref{fig:timecontour}(b), we express $\delta t= c \mathrm{e}^{ib}$ with an infinitesimal constant $c> 0$ and then, the above equation takes the form
 \begin{eqnarray}
\frac{\mathrm{d} \eta_{q}}{\mathrm{d}b}=i\lambda_{E} \sigma_{y} \frac{1}{4 } \eta_{q}.
\end{eqnarray}
The transition occurs from $t_{i}=t_{e}-\epsilon$ to $t_{f}=t_{e}+\epsilon $. Then, when the branch point is present at $y_{0}>0$, the contour $C_{q}$ goes in a clockwise manner, and the angle $b$ goes to $0$ from $-2\pi$. For $y_{0}<0$, $C_{q}$ goes in the opposite direction. Therefore the angle $b$ goes from $0$ to $2\lambda_{b}\pi$ with $\lambda_{b}=-\mathrm{sgn} (y_{0})$, and we obtain the nontrivial solution for the spinor as
 \begin{eqnarray}
\eta_{q}(2\lambda_{b}\pi)
&=&\mathrm{e}^{\rho\frac{\pi i}{2 } \sigma_{y}} \eta_{q}(0),\label{eq:rho}
\end{eqnarray}
with $\rho=\lambda_{E}\lambda_{b} $.
Therefore, the nontrivial solution appears for the band transition (the reflection process).
 
The relation between $\rho$ and physical parameters is obtained more explicitly in the following discussion. 
By the transition amplitude~(\ref{eq:traam}), 
the real part of the instanton action $\mathrm{Re}[iS_{cl}]$ is given as
\begin{eqnarray}
\mathrm{Re}[i S_{cl}] &=&2\int^{y_{0}}_{0}\mathrm{d}y\xi(x(y)+iy)\kappa\Tilde{\kappa} ,
\label{eq:recaction}
\end{eqnarray}
with $t=x(y)+iy$ on $\bar{C}$ . 
$\Tilde{\kappa}=(-1)^{\gamma}$ is given by the number of the transition of the initial Riemann sheet $\gamma$.
 In Fig.~\ref{fig:rs2}(a), we describe a case where the energy transition occurs twice at band edges $k_{A}$ and $k_{B}$ as an example.
The initial time is described by the bold dot on the real time axis, and $\gamma$ increases by one on each transition.
 For the calculation of the semiclassical action, we choose a branch point so that the wavefunction declines, i.e., $\mathrm{Re}[i S_{cl}]<0$. Then, we have relations: 
\begin{eqnarray}
\lambda_{b}=-\mathrm{sgn}(y_{0})=\kappa\Tilde{\kappa}.
\end{eqnarray}
Close to the band edge, we assume the linear kinetic energy $E(k)\sim v (k-k_{e})$, and then we have $\lambda_{E}= \mathrm{sgn}(v) \mathrm{sgn}(y_{0})= -\mathrm{sgn}(v) \lambda_{b} $, because Im$[E(k)]$ does not cross Im$[E(k)]=0$ in the process on $C$.
 Therefore, the sign $\rho$ does not depend on the label of the band $\kappa =\pm 1$, and we have $\rho=\lambda_{E}\lambda_{b}= -\mathrm{sgn}(v)$ for both of the instanton and anti-instanton processes.
Then,
 the transfer matrix is represented as
 \begin{eqnarray}
\mathcal{T}=
\begin{pmatrix}
T ^{*}&\rho\sqrt{1-|T|^2} \\
-\rho \sqrt{1-|T|^2} &T
\end{pmatrix}.\label{eq:transmatgene}
\end{eqnarray}
The magnitude of the reflection coefficient is modified by the conservation law, and the unitarity is satisfied as 
 $\mathcal{T} \mathcal{T}^{\dagger}=1$ by the Berry phase.

\subsection{Adiabatic transition between the valley}
\begin{figure}[htbp]
 \begin{center}
 \includegraphics[width=90mm]{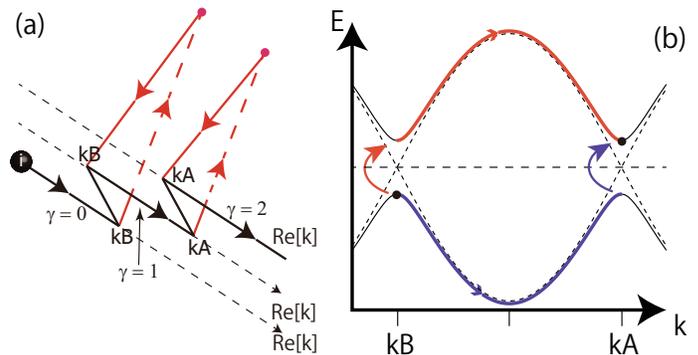}
 \caption{
 (a) Schematic of the transition process between the Riemann sheets. The arrowed solid line describes the transition process of the particle with the force $F>0$, when the transition occurs twice.
The three parallel lines are the real wavenumber (time) axis, and the adiabatic transition occurs there.
 The bold dot on the real wavenumber axis is the initial time of the particle.
 The nonadiabatic transition occurs on $k_{B}$ and $k_{A}$, and the two small dots are the branch points.
 By the transition at the band edge, the number of the Riemann sheets increases by one: $\gamma\to\gamma +1$.
 (b) Schematic of the dispersion of the two-state system with a valley between the two band edges.
 The solid curves show the upper and lower bands $\pm\sqrt{E(k)^2+\Delta^2}$.
Dots show the wavenumbers of the initial and final states.
The dotted curves show the energy band $\pm E(k)$ without interaction ($\Delta = 0$ in the Hamiltonian(\ref{eq:Hamilt})).
Bold arrowed curves show the transition processes, and the interference occurs at the final states on $k_{A}$.
}
\label{fig:rs2}
\end{center}\end{figure}

We consider the adiabatic transition to the band edge $k_{A}$ from one of the nearest-neighbor band edge $k_{B}(<k_{A})$ with $ k_{A}-k_{B}\leq \pi$
 as schematically shown in Fig.~\ref{fig:rs2}(b). 
The solid curves are dispersion $\pm\sqrt{E(k)^2+\Delta^2}$ by the Hamiltonian (\ref{eq:Hamilt}). 
The periodic kinetic energy $\pm E(k)$ is shown by the dotted curves with $E(k)=0$ at $k=k_{A,B}$.
In the vicinity of the band edge $k_{A (B)}$, the system is described as the 1D massive Dirac Hamiltonian by the $k\cdot p$ approximation,
 \begin{eqnarray}
H_{k}\sim H_{A(B)}=(-)v (k-k_{A(B)})\sigma_{z} +\Delta\sigma_{x}.
\end{eqnarray} 
Then, the signs of the Dirac velocity of $H_{A}$ and $H_{B}$ are opposite each other.
According to the previous section, we have the sign of the scattering matrix $\rho_{A}=-\rho_{B}=-\mathrm{sign}(v)$.

We assume that the incident particle goes along the band from $t=t_{B}=k_{B}/F$. At $t_{A}=k_{A}/F$, the particle is on the other band edge $k_{A}$ because $F>0$. The scattering matrix for the adiabatic transition is represented as
  \begin{eqnarray}
\mathcal{U}=
\begin{pmatrix}
\mathrm{e}^{i \frac{1}{F}\theta_{a} }&0\\
0&\mathrm{e}^{-i\frac{1}{F} \theta_{a} }
\end{pmatrix},
\end{eqnarray}
with $ \theta_{a}= \int^{k_{A}}_{k_{B}}\mathrm{d}k~ \xi(k)$.
We note that, the adiabatic transition occurs after the transition of the Riemann sheet, and therefore the sign of the energy is inverted compared with the energy on the original Riemann sheet (Fig.~\ref{fig:rs2}(a)).
Here, we label the scattering matrix on the band edges $k_{A(B)}$ with $\mathcal{T}_{A(B)}$, and then the scattering matrix of the double-passage process has the form
  \begin{eqnarray}
\mathcal{T}= \mathcal{T}_{A} \mathcal{U}\mathcal{T}_{B}.\label{eq:SM}
\end{eqnarray}
 For ideal situations, the adiabatic transition from $k_{A}$ can occur since $k$ is periodic, e.g., the transition from $k_{A}$ to $k_{B}+2\pi$.
In addition, there are materials with band edges more than two in energy spectra.
 Then, the transfer matrix is expressed as
$\mathcal{T}= \mathcal{T}_{N} \mathcal{U}_{N-1} \mathcal{T}_{N-1}\dots\mathcal{U}_{1} \mathcal{T}_{1}$. where $N$ is the number of times that the particle passes through band edges.
In this study, we consider the adiabatic transition between the nearest-neighbor band edges by assuming the coherence length is small, i.e., $N=2$.

\section{The tunneling phenomena in the ladder system}\label{sec:sshl}
Having established the scattering matrix of the LZ problem with the valley, we study the transition phenomenon of a ladder system composed of two quantum wires. 
In the ladder model, an interaction gives a gap, and two band edges appear in the Brillouin zone.
 The hybridization between the quantum wires affects the band structure,
 and the imaginary part of the transition action appears.
Then, we show that the imaginary part evokes the oscillation of the transition probability by varying physical parameters.


 \begin{figure}[htbp]
 \begin{center}
 \includegraphics[width=85mm]{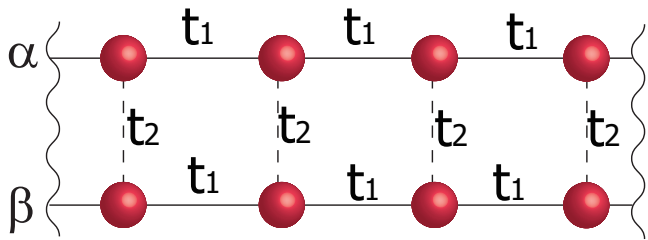}
 \caption{Schematic illustration of the ladder system.
The quantum wires labeled with $\alpha$ and $\beta$ are composed of atoms (sphere) connected by the solid lines.
 The hybridization between the two wires is described by the dotted lines.
Beside these lines, the hopping integral and the magnitude of the hybridization are shown as $t_{1}$ and $ t_{2} $, respectively.
}
\label{fig:tllm}
\end{center}\end{figure}
\subsection{Model}
\begin{figure}[htbp]
 \begin{center}
 \includegraphics[width=85mm]{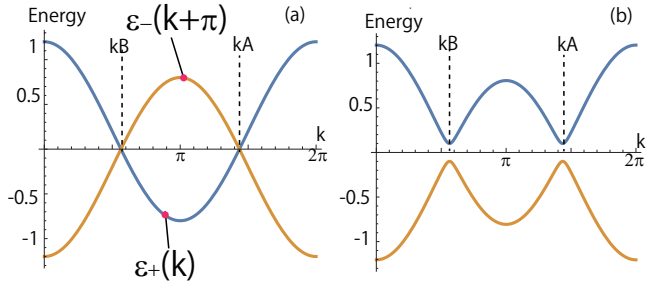}
 \caption{Energy dispersions of the ladder system without(a)/with(b) the correlation in units of $t_{1}$ for $ \frac{t_{2}}{t_{1}} =0.2$.
In (a),  $\epsilon_{+}(k)=E(k)$ and $\epsilon_{-}(k+\pi)=-E(k)$ (Eq.~(\ref{eq:kine})) are shown. 
The Fermi energy is set to be zero, and the degeneracy appears on the Fermi energy at $k_{A,B}$ shown by the dotted vertical lines.
In (b), the gap appears by the correlation $\frac{\Delta}{v}=0.1$, and we used $ \frac{t_{2}}{t_{1}} =0.2$.
The band edges appear at $k_{A,B}$.}
\label{fig:disp}
\end{center}\end{figure}

We employ a ladder lattice system with a constant gap\cite{Ladder}.
We first consider the two wires interacting with each other via the hybridization . 
In the absence of the external field, the Hamiltonian $H_{0}$ is given as
 \begin{eqnarray}
H_{0}&=&t_{1}\sum_{\substack{\mu=\alpha,\beta\\\langle i,j\rangle}} c_{\mu i}^{\dagger}c_{\mu j }+ t_{2} \sum_{ i} ( c_{\alpha i}^{\dagger}c_{\beta i}+ c_{\alpha i}^{\dagger}c_{\beta i})
\end{eqnarray}
The first term is the nearest-neighbor hopping Hamiltonian of the respective quantum wires without interaction. 
$c_{\mu i}$ ($c_{\mu i}^{\dagger}$) is the annihilation (creation) operator of the electron on site $i$ of the quantum wire labeled with $\mu=\alpha,\beta$, and
$t_{1}$ is the nearest-neighbor hopping integral (Fig.~\ref{fig:tllm}).
The second term represents the hybridization of electronic states on the two wires with the magnitude $ t_{2}\geq 0$. We assume $t_{2}\leq t_{1}$ in this study, since the degeneracy is lost for $t_{2}>t_{1}$.
$H_{0}$ is diagonalized in $k$-space as $H_{0k}=\mathrm{diag}(\epsilon_{+},\epsilon_{-})$,
where $\epsilon_{+}(k)= E(k)$ and $\epsilon_{-}=E(k)-2 t_{2} $ with
 \begin{eqnarray}
E(k)=t_{1}\cos k+ t_{2}. \label{eq:kine}
\end{eqnarray}
The Fermi energy is set to be zero, and the lattice constant is expressed as $a=1$ for simplicity.
To focus on the tunneling  phenomena, in the half-hilled model, we introduce a correlation between the upper and lower energy bands by using the mean field approximation on the analogy of the charge-density ordering or the Coulomb (Hubbard)
interaction~\cite{Su79,Su80,Ladder,Giamarchi}.
As shown in Fig.~\ref{fig:disp}(a), the energy spectra  $\epsilon_{+}(k)$ and $\epsilon_{-}(k+\pi))=-E(k)$ are degenerate on the Fermi energy, and then the energy repulsion occurs by the scattering with the wavenumber $\pi$.
The Hamiltonian with the correlation $\Delta$ is given as
\begin{eqnarray}
H
=\begin{pmatrix}
E(k) & \Delta \\
\Delta & -E(k) \\
\end{pmatrix}.
\end{eqnarray}
The eigenvalues are $\xi_{\pm}=\pm \xi_{k}$ with $\xi_{k}= \sqrt{E(k)^2+\Delta^2} $, and the system is gapped (Fig.~\ref{fig:disp}(b)).
In the Brillouin zone, there are two band edges at $k_{A}=\pi+ q_{0} $ and $k_{B}=-q_{0}+\pi$ for $0\leq k\leq 2\pi$,
 where we defined $q_{0}\equiv \arccos \left(\frac{ t_{2} }{t_{1}}\right)$ with $0\leq \arccos \left(\chi\right)< \pi$.
Both the band edges change with the hybridization $ t_{2} $.

By using the wavenumber on the band edge $k_{A(B)}$, $E(k)$ has the form,
\begin{eqnarray}
E_{A(B)}(k)&=& (-)\sqrt{t_{1}^2- t_{2}^2}\sin (k-k_{A(B)})\nonumber\\
&\ &+t_{2} (1-\cos (k-k_{A(B)})).\label{eq:DRM}
\end{eqnarray}
 Then, in the vicinity of the band edges, we obtain the Dirac model by the linear approximation with the group velocity $ \sqrt{t_{1}^2- t_{2}^2}$, when $t_{2}>0$. According to the LZ formula, transition probability is given as $T_{\mathrm{LZ}}^2$, with the amplitude
\begin{eqnarray}
T_{\mathrm{LZ}}&=&\mathrm{e}^{-\frac{\theta_{LZ}}{F} },\nonumber\\
\theta_{LZ}&=&\frac{\pi\Delta^2}{ 2\sqrt{t_{1}^2- t_{2}^2}}.\label{eq:LZP}
\end{eqnarray}
 $\theta_{LZ}$ increases with $t_{2}$, and thus, the transition probability decreases when the hybridization occurs.


\subsection{Transfer matrix of the ladder system}
Here, we obtain the instanton and adiabatic actions to construct the scattering matrix by using Eq.~(\ref{eq:SM}).
Similar to the Dirac model, for the 1D chain with $E(k)\sim \sin k$, the transition process $\xi\to-\xi$ is performed by the purely imaginary time procedure i.e., $t\to i\tau$ with $\tau\in$ Re, since $E(k)$ and $k$ become purely imaginary by the change\cite{Takahashi17}.
Then, the branch cut is parallel to the imaginary time axis.
In the present case, this symmetry of the transformation between $E(k)$ and $k$ is broken when the mass term $ t_{2} (1-\cos k)$ (Eq.~\ref{eq:DRM}) is taken into account; $E(k)$ is not purely imaginary on the imaginary time axis.
Then, the branch cut is curved on the complex plane, and hence the branch points shift in the direction of the real time axis.

The branch cut $\bar{C}$ is given by the constraint Re$[E(k=x+iy)]=$ Re$[t_{1}\cos x\cosh y-it_{1}\sin x\sinh y+ t_{2} ]=0$, and it has the form
 \begin{eqnarray}
 \cos x+\frac{ t_{2} }{t_{1}\cosh y}=0.
\end{eqnarray}
Then, for $0\leq x\leq 2\pi$, the solutions appear in $\frac{\pi}{2}\leq x\leq \pi$ and $\pi\leq x\leq \frac{3\pi}{2}$, and we have $\bar{C}$ on the complex plane:
 \begin{eqnarray}
x_{A(B)}(y)=\pi +\mu_{A(B)} \arccos\left[\frac{ t_{2} }{t_{1}\cosh y}\right], \label{eq:branchcontour}
\end{eqnarray}
 where the sign $\mu_{A(B)}=-1(+1)$ corresponds to $\bar{C}_{A(B)}$ across the band edge $x_{A(B)}(y=0)=k_{A(B)}$. 
Then, we consider the action by the integral along the branch cut $(x_{A(B)}(y), y)$ from the band edge $k_{A(B)}$.
The branch points $y_{0}$ are given by 
 \begin{eqnarray}
 y_{0}=\pm y_{b}= \pm\mathrm{arcsinh}\sqrt{\frac{K^2}{2t_{1}^2} +\sqrt{ \frac{ 4t_{1}^2\Delta^2+ K^4}{4t_{1}^4} } },
\end{eqnarray}
with $K^2=-t_{1}^2+\Delta^2+ t_{2}^2$.
The branch cut $\bar{C}_{A}$ with branch points for $x\geq \pi$ is shown in Figs.~\ref{fig:bpbc} for $\frac{\Delta}{t_{1}}=0.4$(a) and for $\frac{\Delta}{t_{1}}=2$(b).
 The contour of $\bar{C}_{B}$ is symmetry to $\bar{C}_{A}$ with respect to $x=\pi$ according to the relation~(\ref{eq:branchcontour}).
As shown in Figs.~\ref{fig:bpbc}, $\bar{C}_{A}$ is deformed by $t_{2}$.
Due to the shift of the branch point in the direction of the real time axis, the transition cannot be described by a contour along the imaginary time axis from the band edge unlike the Dirac and the 1D chain models. 

By the previous section, the transition action along the contour $C_{A(B)}$ is given as
\begin{eqnarray}
iS_{A(B)}(\kappa\tilde{\kappa})&=&-\frac{i \kappa\Tilde{\kappa}}{F}\int_{C_{A(B)}}\mathrm{d}z~\xi(z)\nonumber\\
&=&
-\frac{1}{F}(\theta_{ R }+ i \mu_{A(B)} \kappa\tilde{\kappa}\theta_{ I } ),
\label{eq:insacl}\\
\theta_{ R }&=& 2\int_{0}^{y_{b}}\mathrm{d}y ~\xi,\label{eq:thetR}\\
 \theta_{ I } &=&\frac{2t_{2}}{t_{1} } \int_{0}^{y_{b}}\mathrm{d}y ~ \frac{ \xi \tanh y}{\sqrt{\cosh^2 y-\left(\frac{t_{2}}{t_{1}}\right)^2}}.\label{eq:thetI}
\end{eqnarray}
We note that the transition amplitude becomes imaginary due to $ \theta_{ I }$, and this comes from the shift of the branch point from the band edge in the direction of the real time axis. 
$C_{A(B)}$ denotes the contour of the transition process along $\bar{C}_{A(B)}$ (Eq.~(\ref{eq:branchcontour})). In Fig.~\ref{fig:bpbc}(a), we show an example of the instanton process $C_{A} $ by the arrowed curve for $t_{2}=0.75$.

Because $F>0$, the initial state is on the band edge on $k_{B}$ with $\Tilde{\kappa}=1$ and $\mu_{B}=-1$ (Eq.~(\ref{eq:branchcontour})).
Then, we obtain $\Tilde{\kappa}\mu_{B}=\Tilde{\kappa}\mu_{A}=-1$ since the sign of  $\Tilde{\kappa}$ is changed by the transition, i.e.,
$iS_{A}(\kappa )=iS_{B}(\kappa )$.
According to Eqs.~(\ref{eq:gentmat}), (\ref{eq:traam}), and (\ref{eq:transmatgene}), the transfer matrices has the form
 \begin{eqnarray}
 \mathcal{T}_{B}&=&
\begin{pmatrix}
T^{*} &R \\
 -R &T
\end{pmatrix}\label{eq:transfB}
\end{eqnarray}
and $
 \mathcal{T}_{A}=
 \mathcal{T}_{B}^{t} $, where the sign of the Dirac velocity is given by Eq.~(\ref{eq:DRM}).
$T=\mathrm{e}^{-\frac{1}{F}(\theta_{ R } +i \theta_{ I })}$ gives the transition probability on the respective band edge is represented as 
\begin{eqnarray}
|T|^2=\mathrm{e}^{-2\frac{\theta_{ R }}{F}},\label{eq:SLP}
\end{eqnarray}
and the reflection coefficient is given as $R=\sqrt{1-|T|^2}$.
Thus, by Eq.~(\ref{eq:SM}),
 we have the transfer matrix for the double-passage transition as
 \begin{eqnarray}
\mathcal{T}&=&
\begin{pmatrix}
R^2\mathrm{e}^{-i\frac{1}{F}\theta_{a}} +T^{*2}\mathrm{e}^{i\frac{1}{F} \theta_{a}}&R(T\mathrm{e}^{-i\frac{1}{F}\theta_{a}}-T^{*}\mathrm{e}^{i\frac{1}{F}\theta_{a}})\\
-R(T\mathrm{e}^{-i\frac{1}{F}\theta_{a}}-T^{*}\mathrm{e}^{i\frac{1}{F}\theta_{a}})&R^2\mathrm{e}^{i\frac{1}{F}\theta_{a}}+T^{*2}\mathrm{e}^{-i\frac{1}{F}\theta_{a}}
\end{pmatrix},\nonumber\\
\label{eq:ttm}
\end{eqnarray}
where the adiabatic transition is given as $ \theta_{a}=2\int_{0}^{q_{0}  }\mathrm{d}k~\sqrt{(t_{1}\cos k-t_{2})^2+\Delta^2 }$.
The total transition amplitude is given $ T_{d}\equiv \mathcal{T}_{12}$ by 
the off-diagonal element of the transfer matrix as
 \begin{eqnarray}
T_{d} 
&=&2i\sqrt{1-|T|^2}|T|\sin\left(\frac{\theta_{I}+\theta_{a}}{F} \right),
\label{eq:dptrans}
\end{eqnarray}
 and the transition probability is given as $|T_{d}|^2$.

When we take into account the Stokes phase\cite{Kayanuma97,Wubs05,Nori10}, $\varphi_{\mathrm{st}}$, 
the reflection coefficient (\ref{eq:transfB}) has the phase, $R\to R\mathrm{e}^{\pm i\varphi_{\mathrm{st}}+i\pi}$: 
 \begin{eqnarray}
 \mathcal{T}_{B}&\to&
\begin{pmatrix}
T^{*} & R\mathrm{e}^{ i\varphi_{\mathrm{st}}+i\pi} \\
 -R\mathrm{e}^{- i\varphi_{\mathrm{st}}+i\pi} &T
\end{pmatrix}.
\end{eqnarray}
In this case, the shift of the oscillation of probability occurs by the change of the sinusoidal function in Eq.~(\ref{eq:dptrans}), $ \sin\left(\frac{\theta_{I}+\theta_{a}}{F} \right)\to \sin \left(\frac{\theta_{I}+\theta_{a}}{F}+\varphi_{\mathrm{st}}\right) $.
When we neglect the lattice effect, we obtain the amplitude $T \to T_{\mathrm{LZ}}$ with $\theta_{I}\to 0$, 
and the conventional interband transition probability is obtained.
We note that the sinusoidal function comes from the dynamical adiabatic phase $\theta_{a}$ and the imaginary part of the transition amplitude $\theta_{I}$ by the semiclassical description, and the transition amplitude oscillates by the presence of $\theta_{I}$ even when $\theta_{a}=\varphi_{\mathrm{st}}=0$.
For $t_{1}=t_{2}$, we have $\theta_{a}=0$, and the transition occurs at the parabolic bands ~\cite{Suominen99}. In this case, $\theta_{I}$ becomes dominant in the oscillation phase. 

\section{Transition probability for the ladder system}\label{sec:prob}
In the previous section, we have formulated the transfer matrix by using the instanton and adiabatic actions, to compute the total transition amplitude for the ladder system.
 In actual systems, when the relaxation of the quasi particle occurs for the adiabatic transition,
 the interference of the transition amplitude does not occur.
 When the relaxation time of the particle is given as $\Tilde{\tau}$, the condition for the coherent transport between the two band edges 
is given as $ \tau_{d}=\frac{|k_{A}-k_{B}| }{F}=\frac{2}{F} q_{0}  < \Tilde{\tau}$.
In this section, we first study the transition probability with the small relaxation time $ \Tilde{\tau} < \tau_{d}$. 
This situation will occur when the coherence length is small by disorder.
 Next, we discuss the double-passage process for $ \tau_{d}< \Tilde{\tau}$.
 The calculation results show the oscillation of the probability, which is strongly affected by the hybridization.

\subsection{Transition of single-passage process}\label{sec:probad}
\begin{figure}[htbp]
\begin{center}
 \includegraphics[width=90mm]{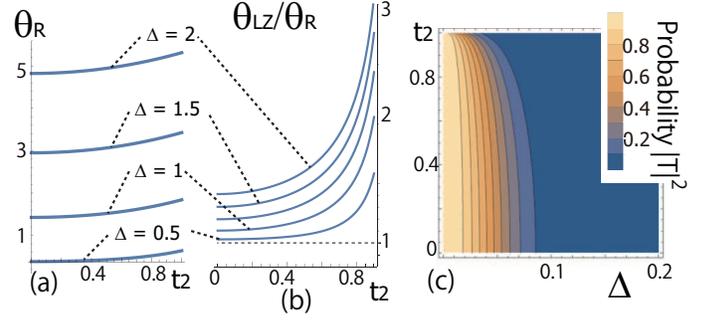}
 \caption{ (a) Magnitude of the function $\theta_{ R }$ by Eq.~(\ref{eq:thetR}) is shown as a function of the hybridization $t_{2}$ for $\Delta=0.5$-$2$ in units of $t_{1}$.
$\theta_{ R }$ increases with $t_{2}$. (b) The difference between the lattice and Dirac models on the transition amplitude is shown by the ratio $\frac{ \theta_{ LZ }}{\theta_{R}}$. The horizontal dotted line shows $\frac{ \theta_{ LZ }}{\theta_{R}}=1$. By the results $\theta_{LZ} >\theta_{ R }$, the tunneling probability by the LZ formula is smaller than that by instanton method of the Bloch states.
 $\theta_{ R }$ and $\frac{\theta_{R}}{ \theta_{ LZ }}$ do not depend on $F$.
(c) Distribution of the tunneling probability of the ladder system $|T_{d}|^2$ is shown in the $\left(\frac{\Delta}{t_{1}}, \frac{t_{2}}{t_{1}}\right)$ plane with $t_{1}=1$. The magnitude of $|T_{d}|^2$ is shown by color. The probability decreases by increasing $\Delta$ or $t_{2}$.
}
\label{fig:ThPr}
\end{center}\end{figure}

When the relaxation time is small  $ \Tilde{\tau} < \tau_{d}$, the transition processes on $k_{A}$ and $k_{B}$ occurs as independent events,
 and the tunneling probability on the respective band edges is given by the real part of the instanton action, i.e., $|T|^2$ (Eq.~(\ref{eq:SLP})).
In Figs.~\ref{fig:ThPr}(a)(b), the calculation results of $\theta_{ R }$ and $\frac{\theta_{ R }}{\theta_{ LZ}}$ are shown. 
These functions do not depend on the external force $F>0$. Namely they are determined by the intrinsic parameter of the material, $t_{1}$, $t_{2}$, and $\Delta$.

In Fig.~\ref{fig:ThPr}(a), we find that $\theta_{ R }$ is an increasing function of $t_{2}$, and therefore the tunneling amplitude decreases by the increase in $t_{2}$. 
 This tendency is similar to the probability by the LZ formula $\theta_{LZ}$ (Eq.~(\ref{eq:LZP})).
 By Fig.~\ref{fig:ThPr}(b), we obtain the difference between the probabilities by the lattice and Dirac systems  $\frac{\theta_{ R }}{\theta_{ LZ}}$ as  a function of $t_{2}$. 
The results give $\theta_{LZ}>\theta_{ R }$, and this means that the transition amplitude for the lattice system is larger than that by the LZ formula.
The Dirac velocity decreases with $t_{2}$, and $\theta_{LZ}$ diverges at $t_{1}=t_{2}$.
 On the other hand, $\theta_{R}$ does not diverge on $t_{2}=t_{1}$, due to the higher order of the periodic energy $E_{A,B}(k)$\cite{Takahashi17},
 and this means that the transition cannot be described by the LZ formula when $t_{2}$ becomes large.
Figure~\ref{fig:ThPr}(c) shows $|T_{d}|^2$ on the $\left(\frac{\Delta}{t_{1}}, \frac{t_{2}}{t_{1}}\right)$ plane, for the external force $\frac{F}{t_{1}}=10^{-3}$.
As shown in (c), $|T|^2$ monotonically decreases with $t_{2}$ or $\Delta$, and this is consistent with the behavior of $\theta_{ R }$ in (a).
 For the case with $ \Tilde{\tau} < \tau_{d}$, when we take into account the transition on the two band edges, the total tunneling probability has the form $\propto |T|^2(1-|T|^2)$.

\begin{figure}[htbp]
\begin{center}
 \includegraphics[width=85mm]{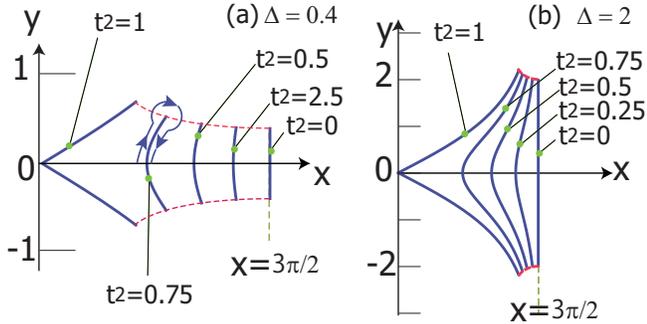}
 \caption{The branch cut ($\bar{C}_{A}$) at $x>\pi$ and branch points for $\Delta=0.4$(a) and for $\Delta=2$(b) with $t_{2}=0-0.75$, in units of $t_{1}$. The solid curves are $\bar{C}_{A}$, and the dotted curves show the track of the branch points by the change of $t_{2}$. 
The distance between the branch point and band edge becomes large with the increase in $t_{2}$ or $\Delta$. In (a), a schematic of the instanton process is shown by the arrowed curve around the branch cut for $t_{2}=0.75$. For the calculation of the transition amplitude, the distance between the branch cut and the contour is infinitesimal.
}
\label{fig:bpbc}
\end{center}\end{figure}

\subsection{Double-passage process}\label{subsec:DDP}
Here, we consider the system with the large relaxation time $ \Tilde{\tau}> \tau_{d}$ and show the calculation results of the tunneling probability~(\ref{eq:dptrans}) of the double-passage process, without the quantum phase to focus on the semiclassical lattice effect, i.e., $\varphi_{\mathrm{st}}=0$.
Then, the oscillation of the probability is shown. 
The observation of the oscillation enables to estimate the transition action, and we will estimate physical parameters by numerical calculations.

\subsubsection{Comparison between $\theta_{I}$ and $\theta_{a}$}
\begin{figure}[htbp]
\begin{center}
 \includegraphics[width=85mm]{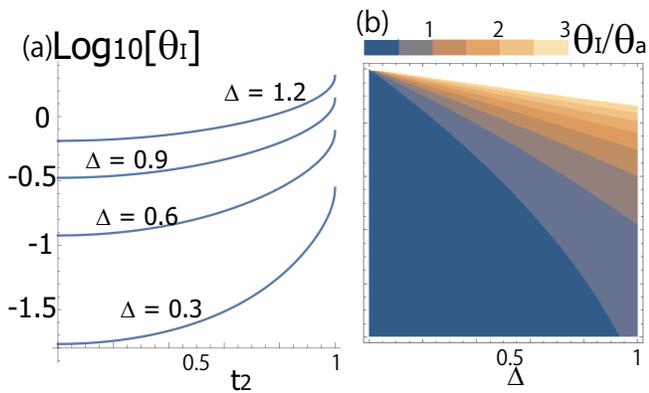}
 \caption{ (a) Magnitude of the imaginary part of the transition amplitude $\theta_{I}$ for $\Delta=0.3-1.2$ as a function of $t_{2}$ in a log scale.
 We see that $\theta_{I}$ increases with the gap $\Delta$ or hybridization $t_{2}$.
(b) The ratio between $|\theta_{I}|$ and the magnitude of the dynamical phase $|\theta_{a}|$, $\left|\frac{\theta_{I}}{\theta_{a}}\right|$ in the $\left(\frac{\Delta}{t_{1}}, \frac{t_{2}}{t_{1}}\right)$ plane. 
The magnitude of the ratio is shown by color, and in the value of the white region is larger than 3.
 $|\theta_{I}|$ is much larger than $|\theta_{a}|$ for large $t_{2}$ since $|\theta_{a}|$ decreases with $t_{2}$, and finally, $\theta_{a}=0$ at $t_{2}=t_{1}$. 
$t_{1}=1$ is used in (a) and (b).}
\label{fig:DPPhase}
\end{center}\end{figure}

We consider the transition phase evoking the oscillation.
In Figs.~\ref{fig:DPPhase} (a) (b), the magnitude of $ \theta_{I}$ is shown as a function of $t_{2}$ in a log scale, $\log_{10}[\theta_{I}]$, for $\Delta=0.3-1.2$, and the ratio $\left|\frac{\theta_{I}}{\Tilde{\theta}}\right|$ is calculated in the $(\frac{\Delta}{t_{1}}, \frac{t_{2}}{t_{1}})$ plane.
 By Fig.~\ref{fig:DPPhase}(a), we find that the oscillation by the lattice effect becomes large when the gap $\Delta$ and the hybridization $t_{2}$ are large. This is consistent with the calculation results of the branch cut (Figs.~\ref{fig:bpbc}), because the shift of the branch point along the real wavenumber becomes large with these parameters. 
By  Fig.~\ref{fig:DPPhase}(b), we find that the lattice effect $\theta_{I}$ is dominant in the phase factor when $t_{2}$ is large, since the adiabatic action $\theta_{a}$ decreases with $t_{2}$ and vanishes at $t_{2}=t_{1}$.
 $\theta_{I}$ is comparable to or larger than $\theta_{a}$ in a wide range. Therefore, the lattice effect on the oscillation of the probability cannot be neglected in the double-passage process, and to obtain the oscillation by the lattice effect, large values of $t_{2}$ are favorable.

\subsubsection{The oscillation by varying physical parameters}
\begin{figure}[htbp]
\begin{center}
 \includegraphics[width=90mm]{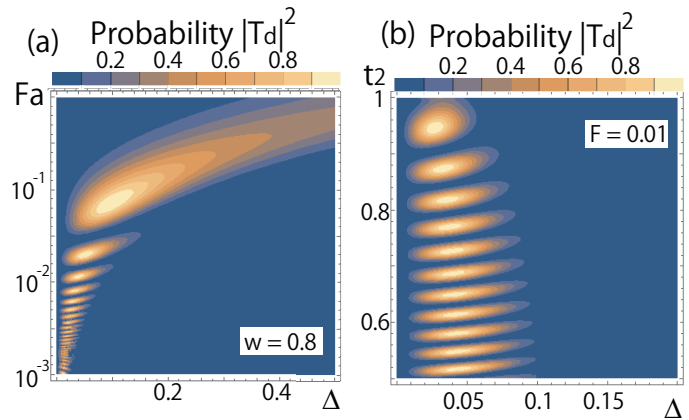}
 \caption{ Contour plots of the transition probability $|T_{d}|^2$ (Eq.~(\ref{eq:dptrans})) of the double-passage problem in the $\left(\frac{\Delta}{t_{1}}, \frac{F }{t_{1}}\right)$ plane for $t_{2}=0.8$ (a), and in the $\left(\frac{\Delta}{t_{1}}, \frac{t_{2}}{t_{1}}\right)$ plane at $F=10^{-2}$(b).
 $t_{1}=1$ is used in (a) and (b). 
$|T_{d}|^2$ oscillates due to the two phase factors by $\theta_{a}$ and $\theta_{I}$. 
(a) shows that the oscillation becomes intense by decreasing $F$, since the actions are proportional to $\frac{1}{F}$.
In (b), the oscillation becomes moderate by increasing with $t_{2}$, since $\theta_{a}$ decreases with $t_{2}$.
}
\label{fig:DPP22}
\end{center}\end{figure}

We show the calculation results of the probability $|T_{d}|^2$ (Eq.~\ref{eq:dptrans}) and find the oscillation by varying the parameters as shown in Figs.~\ref{fig:DPP22} (a) and (b).
For (a), $|T_{d}|^2$ is shown in the $(\frac{\Delta}{t_{1}}, \frac{F}{t_{1}})$ plane with the hybridization $\frac{t_{2}}{t_{1}}=0.8$.
The strong oscillation appears along $F$, and  maximal values of the probability are explicitly seen when $\Delta$ is much larger than $F$.
The oscillation becomes intense for small values of $F$, because the action is proportional to $\frac{1}{F}$.
Figure~\ref{fig:DPP22}(b) shows the oscillation in the $\left(\frac{\Delta}{t_{1}}, \frac{t_{2}}{t_{1}}\right)$ plane, and we use $\frac{F}{t_{1}}=10^{-2}$.
The oscillation becomes moderate by the increase in $t_{2}$, because $\theta_{a}$ decreases and $\theta_{a}=0$ at $t_{2}=t_{1}$.


\subsubsection{Interval of the oscillation}
Maximal values of the probability appear when the derivative of the amplitude (\ref{eq:dptrans}) is zero.
Here, we consider the oscillation by the increase in the force $F$, for the small transition amplitude on the respective band edge , i.e., $\mathrm{e}^{-2\frac{\theta_{R}}{F}}\ll1$.
 In this case, by $\partial_{F}T_{d}=0$, we have 
 \begin{eqnarray}
\cos\left( \frac{\theta_{a}+\theta_{I}}{F}+ \Tilde{\theta}\right)=0,
\end{eqnarray}
where $\Tilde{\theta}$ does not depend on $F$ with $\cos\Tilde{\theta}= \frac{\theta_{I}}{\sqrt{\theta_{I}^2 +4\theta_{R}^2}}$ and $\sin\Tilde{\theta}= \frac{2\theta_{R}}{\sqrt{\theta_{I}^2 +4\theta_{R}^2}}$. 
Then, the interval of the force between the $m$-th and $(m-1)$-th maximal probabilities is given as
\begin{eqnarray}
\frac{1}{F_{m}}-\frac{1}{F_{m-1}} =\frac{\pi}{\theta_{I}+\theta_{a}},\label{eq:intf}
\end{eqnarray}
where $F_{m}$ is the force giving the $m$-th maximal probability.
 The above r.h.s. does not depend on $F$, then the interval of $F^{-1}$ is determined by the material parameters.  
Similarly, equations for the maximal probability can be done by the derivative of the amplitude (\ref{eq:dptrans})
 with respect to $(t_{1},t_{2},\Delta)$, although they become complex form.

%

\subsubsection{Comparison between the Stokes phase and the transition phase of the lattice effect}
\begin{figure}[htbp]
\begin{center}
 \includegraphics[width=80mm]{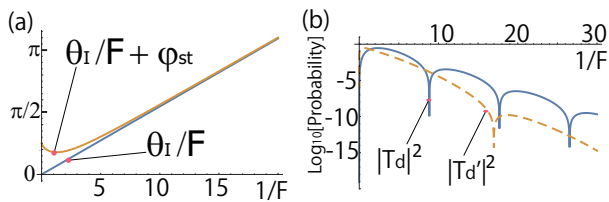}
 \caption{(a) Plot of the transition phases (modulo $2\pi$) as a function of the inverse force $\frac{1}{F}$.
 $\theta_{I}$ (solid curve) is the phase by the nonlinear effect of the lattice, and $\varphi_{\mathrm{st}}$ is the Stokes phase (dotted curve).
For the small force, the contribution from $\varphi_{\mathrm{st}}$ can be neglected.
(b) Plot of the transition probabilities of double-passage problem as a function of $\frac{1}{F}$.
The solid and dotted curves show $|T_{d}|^2$ and $|T_{d}'|^2$, respectively. 
The probability on vertices is zero.
In both of the results(a)(b), $\left(\frac{t_{2}}{t_{1}}, \frac{\Delta}{t_{1}}\right)=(0.9,0.4)$ is used. 
}
\label{fig:stF}
\end{center}\end{figure}

We mention the Stokes phase $\varphi_{\mathrm{st}}=\frac{\pi}{4} +\delta(\log\delta -1  ) +\arg[\Gamma(1-i\delta)]$ as a quantum correction of the double-passage process\cite{Nori10,Kayanuma97,Wubs05}, with $\delta =\frac{\Delta^2}{2F\sqrt{t_{1}^2-t_{2}^2}}$.
In our main calculation results (Sec.~\ref{sec:probad}), we neglected $\varphi_{\mathrm{st}}$ to focus on the oscillation by the lattice effect.
In addition, we have shown the transition probabilities by the Dirac and lattice models, since they are derived by the semiclassical approximation.
Here, we show that the Stokes phase can be neglected when the lattice effect becomes prominent. 
We consider the case with the strong lattice effect in the oscillation; $\theta_{I}$ is dominant compared with  $\theta_{a}$(Fig.~\ref{fig:DPPhase}(b)). 
In Fig~\ref{fig:stF}(a), the phases, $\frac{\theta_{I}}{F}$ and $\frac{\theta_{I}}{F}+\varphi_{\mathrm{st}}$  (modulo $2\pi$) are shown,
 and we find that the contribution to the phase by $\varphi_{\mathrm{st}}$ becomes small for small values of $F$.
To show a comparison between the transition probability in the presence or absence of the lattice effect, we define the probability $|T_{d}'|^2$ conventionally used in the double-passage problem, where
\begin{eqnarray}
T'_{d}=2iT_{\mathrm{LZ}}\sqrt{1-T_{\mathrm{LZ}}^2}\sin \left(\frac{\theta_{a}}{F}+ \varphi_{\mathrm{st}}\right).
\end{eqnarray}
In Fig~\ref{fig:stF}(b), $|T_{d}|^2$ and $|T_{d}'|^2$ are plotted as a function of the inverse force $\frac{1}{F}$ at $\left(\frac{t_{2}}{t_{1}}, \frac{\Delta}{t_{1}}\right)=(0.9,0.4)$.
 Then, the strong fluctuation of the probability is seen by $|T_{d}|^2$.
By the above results, we find that the interval between the maximal probabilities is different for small values of $F$, and we can neglect $\varphi_{\mathrm{st}}$.
For materials, we have $F<\Delta$, because the system is not stable when the force becomes $F\sim\Delta$, and hence the lattice effect should be taken into account for the oscillation phenomenon.

\section{Discussion and summary}\label{sec:sum}
 In this work, we focus on the lattice effect of the ladder system by the semiclassical description.
As for the transition at the respective band edges, the lattice effect evokes the enhancement of the tunneling probability compared with that of the LZ formula, and this is similar to the previous work. 
By our results, the oscillation appears by an analysis of the semiclassical regime.
Then, the lattice effect is expected to be robust, since it comes from the semiclassical saddle point, unlike quantum corrections.
By our results, the intense oscillation is seen for small $\frac{F}{t_{1}}$ or $\frac{t_{2}}{t_{1}}$, and we have $\varphi_{\mathrm{st}}\sim 0 $ in these parameter regions.
 Although we may obtain small values of the transition probability with the large lattice effect, the tunneling signal is enhanced by increasing the number of the ladder systems.
 This will be done by employing the large cross-section of the insulator, in a simple way.

The oscillation is expected to be applied to current controls.
The physical parameters of the ladder system can be tuned in several ways.
 The hybridization $t_{2}$ can be controlled by the gate voltage or pressure\cite{RibeiroRM09}, perpendicular to the direction of the wires; by applying the potential difference $V_{g}$ between the two wires, the hybridization reduces to $t_{2}\to\sqrt{t_{2}^2+V_{g}^2}$.
Therefore, the observation in this paper will be done in optical lattice or solid-state materials.
 Our findings of the oscillation is expected for various tunneling phenomena, since the imaginary part of the transition amplitude comes from the nonlinearity of the energy spectra. 
When the curvature of the energy spectra is large, the transition amplitude will obtain the imaginary part,
 and the oscillation of the probability can be seen when the particle encounters the anti-crossing more than twice.


In summary, we study the tunneling phenomenon in a ladder system by using the instanton method,
 and find the oscillation of the tunneling probability due to the periodic nature of the lattice system when the relaxation time is large.
 We first construct the transfer matrix of the instanton process for two-state systems with a constant gap.
Then, the transition probability is calculated in a simple two-leg ladder model, and we find that the nonlinearity of the energy spectra contributes to the transition phase. 
The nonlinearity is given by the periodicity of the lattice system,
 and this effect on the oscillation becomes strong when the hybridization between the ladder is large.
In addition, when the force becomes small, the oscillation is enhanced, and therefore, the oscillation becomes prominent when the probability is small.

\end{document}